\documentstyle[preprint,aps]{revtex}
\begin{document}

\title{Enhancement of low-mass dileptons in heavy ion collisions}
\bigskip
\author{G. Q. Li$^a$, C. M. Ko$^a$, and G. E. Brown$^b$}
\address{$^a$Cyclotron Institute and Physics Department,\\
Texas A\&M University, College Station, Texas 77843\\
$^b$Department of Physics, State University of New York,
Stony Brook, NY 11794}

\maketitle

\begin{abstract}
Using a relativistic transport model for the expansion stage of
S+Au collisions at 200 GeV/nucleon, we show that the recently observed
enhancement of low-mass dileptons by the CERES collaboration
can be explained by  the decrease of vector meson masses in hot
and dense hadronic matter.
\end{abstract}


\medskip

Based on the broken scale invariance of QCD, Brown and Rho have shown that
the masses of non-strange vector mesons should be reduced in dense matter
\cite{brown91,brown93}. This is supported by studies using the QCD
sum rules \cite{hatsu92}.  Experimentally, this can be verified
by measuring the dileptons produced from heavy ion collisions.
Since dileptons are not subject to the
strong final-state interactions associated with hadronic observables,
they are the most promising probe of
the properties of hot dense matter formed in the initial stage of
high energy heavy ion collisions
\cite{shur78,chin82,kaja86,gale87,xia88,xia90,asak93,asak94}.
Indeed, our recent study using the relativistic transport model
has shown that in heavy ion collisions
at SIS/GSI energies significant differences exist between
the dilepton spectra with and without medium
modifications of the masses and widths of vector mesons \cite{li95}.
In particular, we have found that the rho peak from pion-pion
annihilation shifts to around 550 MeV, and its height increases by
about a factor of four.

Dileptons have already been measured at Bevalac/LBL by the DLS
collaboration \cite{dls88} in heavy-ion collisions at incident energy
around 1 GeV/nucleon.  Theoretical studies have shown that
the observed dileptons with invariant masses above about 450 MeV
are mainly from pion-pion annihilation \cite{xiong90,wolf93}.
Unfortunately, statistics are not good enough in the Bevalac experiments
to give definite information on the in-medium vector meson properties.
However, similar experiments with vastly improved statistics
has been planned at SIS/GSI by the HADES collaboration
\cite{hades94}.

A calculation similar to that of Ref. \cite{li95} has been carried out
in the RQMD \cite{hoff94}.  Using the in-medium rho meson mass from
the QCD sum rules, a shift in the rho meson peak in the dilepton
invariant mass spectrum is also seen in heavy ion collisions at AGS/BNL
energies.  Because of higher incident energies than at SIS/GSI,
about 75\%
of the rho mesons in this study are produced from reactions other
than pion-pion annihilation.  Unfortunately, no experiments at the
AGS/BNL have been designed to measure dileptons to verify the theoretical
predictions.

For heavy ion collisions at the SPS/CERN energies, hot and dense
matter is also formed in the initial stage of the collisions.  We expect
that medium effects will also lead to a shift in the vector meson peaks
in the dilepton invariant mass spectra. Experiments from both the HELIOS-3
\cite{helios95} and the CERES \cite{ceres95} collaboration have shown
that there is an excess of dileptons over those known and expected
sources which can not be explained by uncertainties and errors of the
normalization procedures.  In particular, in the CERES experiment on
central S+Au collisions at 200 GeV/nucleon, a significant enhancement
of dileptons with invariant masses between 250 MeV to 1 GeV over that
from the proton-nucleus collision has been found.  In this Letter, we
shall show that the modification of vector mass properties in medium
can explain the enhancement of dileptons in this mass region.

To study this quantitatively, we generalize
the relativistic transport model \cite{ko87,li94}, which is
based on the nonlinear $\sigma$-$\omega$ model for the nuclear matter
\cite{qhd}, to describe the expansion of a hot dense fire-cylinder that
is expected to be formed in the S+Au collisions
at 200 GeV/nucleon.  The initial conditions of this fire-cylinder are
determined by fitting the observed pion and proton rapidity distributions
and transverse mass spectra after the full dynamical evolution of
the system.  Specifically, they are fixed by the data from
the NA44 collaboration \cite{na44} on central S+Pb collisions
at 200 GeV/nucleon, which is very similar to the S+Au collisions in
the CERES collaboration.

To describe reasonably the proton rapidity
distribution, we find that the baryon number in the
initial fire-cylinder is about 72. We include all baryon
resonances with masses below 1720 MeV.
We also include the lowest-lying hyperons, i.e., lambda ($\Lambda$)
and sigma ($\Sigma$).
Assuming that there are 32 baryons from the projectile and 40 from the
target, the center-of-mass rapidity of the fire-cylinder with respect
to the laboratory frame is then 2.63, which is close to the prediction
from the RQMD simulation \cite{hoff94}.
Initially, these baryons are distributed
in a fire-cylinder whose cross section is taken to be about 40 fm$^{2}$,
similar to the geometrical cross section of the projectile nucleus. If we
further assume that the initial baryon density is about 3.5$\rho _0$
($\rho _0 = 0.16$ fm$^{-3}$), then the longitudinal length $2z_L$ of the
fire-cylinder is found to be 3.2 fm.
For mesons, we include pions, rhos, and omegas, as well as kaons
and antikaons. Their intial abundance is determined by requiring that
the final pion number agrees with the measured value and will be specified
later. Furthermore, both baryons and
mesons are distributed uniformly within the fire-cylinder.

The initial transverse momentum distributions of these particles are
assumed to be given by a thermal distribution.   We find that an initial
temperature
of about 185 MeV is needed to reproduce the observed slopes of the
transverse momentum spectra for both protons and pions. Their longitudinal
momentum distributions are determined by imposing a rapidity field as in
the hydrodynamical model \cite{bjor83,bolz92}. Specifically, we assume
that the rapidity of a particle in the fire-cylinder frame
is correlated to its longitudinal position via $y=1.2\,\sinh (z/z_L)$.
Thus, particles at the surface of the cylinder move faster than those
in the interior.

The fire-cylinder is
then evolved as in the usual transport model for heavy-ion collisions
\cite{ko87,li94}. For baryon-baryon interactions, we include both
elastic and inelastic scattering for nucleons and deltas (1232), while
for higher resonances we consider only elastic scattering with cross sections
taken to be the same as that for NN scattering at the same center-of-mass
energy. The meson-baryon interactions are formulated through resonance
formations and decays. For meson-meson interactions, we include both
pion-pion annihilation to a rho meson and the rho meson decay. This
process is mainly responsible for the production of dileptons with
invariant masses from 2$m_\pi$ to $m_\rho$. The decay of an omega
into three pions is also explicitly included, but the reverse process of
omega formation from three pions is neglected. As the omega has a very
small decay width ($\approx 8$ MeV), this is expected to have negligible
effects on dilepton production.

In the calculation, dileptons are all produced from the decay of
rho and omega mesons, i.e., $\rho ^0\rightarrow e^+e^-$, and $\omega
\rightarrow e^+e^-$. The contribution of pion-pion annihilation
is treated as a two-step process with explicit rho meson formation,
propagation and decay, i.e., $\pi ^+\pi ^- \rightarrow \rho ^0\rightarrow
e^+e^-$. As shown in Ref. \cite{li95}, to study properly medium
effects, it is essential to treat the intermediate vector meson
explicitly in meson-meson annihilation.

For a pair of pions with a total invariant mass $M$, a rho meson
of this mass is formed with a cross section given by the Breit-Wigner
form \cite{li95}
\begin{eqnarray}
\sigma _{\pi ^+\pi ^-\rightarrow \rho ^0} = {12\pi\over k^2}
{(m_\rho \Gamma _\rho )^2\over (M^2-m_\rho ^2)^2+(m_\rho \Gamma _\rho )^2},
\end{eqnarray}
where $k$ is the pion momentum in the center-of-mass frame of the
rho meson.  In the above, $m_\rho$ and $\Gamma_\rho$ are the centroid
and width of the rho meson.  When the medium modification of the
rho-meson mass is included, $m_\rho$ and $\Gamma _\rho$ are replaced
by $m_\rho^*$ and $\Gamma _\rho^*$, respectively \cite{li95}.

The decay width of the rho meson into a dilepton is proportional
to its mass \cite{bhaduri88}, i.e., $\Gamma _{\rho ^0\rightarrow
e^+e^-} (M) =\lambda _\rho M,$ where $\lambda _\rho = ~8.814\times
10^{-6}$ is determined from the observed width at $m_\rho \approx$ 768 MeV.
Similarly,  for the omega decay width, we have
$\Gamma_{\omega\rightarrow e^+e^-} (M) =\lambda _\omega M,$
with $\lambda _\omega = ~ 0.767 \times 10 ^{-6}$.

To make a quantitative comparison with the experimental data, we need to
include the experimental acceptance and resolution. The former is
taken into account by first transforming the dilepton momentum and energy
from the fire-cylinder frame to the laboratory frame, using the rapidity
determined previously, and then including only those dileptons with
transverse momentum $p_t> 0.2$ GeV and pseudo-rapidity
$2.1< \eta < 2.65$. The effect from the cut-off
in the opening angle $\theta _{ee}>35 $ mrad is estimated to be very small
(on the level of 10\%) and has been neglected. Following Ref. \cite{asak94},
the experimental resolution is included by folding the original dilepton
mass spectrum with a normalized smearing function of the Gaussian form
with a variance $\sigma\approx 15$ MeV, similar to the mass resolution
in the CERES experiment.

We first consider the case in which both rho and omega mesons
are treated as free particles.  We assume that pions, rhos, and omegas
are in chemical equilibrium, i.e,
$\mu _\rho =2\mu _\pi$ and $\mu _\omega =3\mu _\pi$.
In order to reproduce the observed pion rapidity distribution and
spectrum, we find that the pion chemical potential is about 135 MeV at
an initial temperature of 185 MeV. The initial pion, rho and omega
numbers are then determined to be 95, 57, and 39, respectively.
At this temperature,
we find that nucleons and deltas (1232) account for about 60\% of all
baryons, while for higher resonances each contributes about 1-3\%.

The final proton and pion transverse momentum spectra and rapidity
distributions are shown in Fig. 1 by the dotted histograms.
They are seen to agree reasonably with the preliminary data
from the NA44 collaboration for central S+Pb collisions at 200 GeV/nucleon
shown in Fig. 1 by solid circles.
We note that initially both protons and pions
are assumed to have the same temperature.  The final pion spectrum is,
however, steeper than the proton spectrum, and this is mainly
due to the transverse expansion of the system.

There are three sources for dilepton production in our model,
i.e., the decay of primary rho mesons (those already exist in the
initial fire-cylinder), the decay of omega mesons, and the decay of
rho mesons formed from pion-pion annihilation during the evolution of
the system. The latter one is usually identified as the contribution
from pion-pion annihilation.  As expected,  a strong omega peak is
seen around 780 MeV, which becomes flatter after folding with the
experimental resolution. Also, we see a broad rho peak around 770 MeV
due to decays of primary rho mesons. If there were no pion-pion
annihilation, then the dilepton yield would decrease substantially
once its mass is below $m_{\rho,\omega}$. Pion-pion annihilation, on
the other hand, builds up the dilepton strength between
2$m_\pi$ and $m_\rho$, as already shown in heavy ion collisions
at Bevalac energies \cite{xiong90,wolf93}. Therefore, the CERES data
clearly indicate the importance of pion-pion annihilation, as noted in
Ref. \cite{ceres95}. As shown in Fig. 2 by the dotted histogram, even
with the inclusion of pion-pion annihilation, the theoretical dilepton
yield disagrees  with the experimental data.  The former is about a
factor of 3-5 too low between 2$m_\pi$ and 550 MeV, and about a factor
of 3 too high  around $m_{\rho ,\omega}$.

The effects of the medium modification
of vector mesons can be consistently included in our model
by extending the non-linear
$\sigma$-$\omega$ model used in our previous studies \cite{li95}
and in the above calculation
to include the explicit coupling  of vector mesons to the scalar field
using the idea of the quark-meson coupling model of Ref.
\cite{thomas94}. The generalized scalar field then satisfies the
following self-consistent conditions

\begin{eqnarray}
m^*_q&=&m_q-{1\over 3}g_{\sigma}\langle\sigma \rangle,\nonumber\\
m_\sigma ^2 \langle\sigma\rangle +b \langle\sigma \rangle^2 +c\langle\sigma
\rangle^3
&=&g_{\sigma} \rho _{sN} + {2\over 3}g_{\sigma} \rho _{s\rho} +
{2\over 3}g_{\sigma} \rho _{s\omega},
\end{eqnarray}
where the constituent quark mass is denoted by $m_q$, and
$\rho _{sN}, ~\rho_{s\rho},$ and $\rho _{s\omega}$ are the
`scalar' densities of the nucleon, the rho meson, and the omega meson,
respectively.  The parameters $g_{\sigma}$, $b$ and $c$ are taken from
Ref. \cite{li94} that correspond to a soft nuclear equation of state with
a nucleon Dirac mass $m_N^*\approx 0.83\, m_N$ and a compressibility
$K\approx 200$ MeV at normal nuclear density. The nucleon, rho-meson and
omega-meson masses are then given by
$m^*_N\approx 3 m_q^*$ and $m^*_\rho\approx m^*_\omega \approx
2 m_q^*$ according to the constituent quark model.
In this model, the scalar field energy is large in hot dense matter
when hadron masses are reduced.  As the system expands and cools, the
field energy decreases and is converted back to hadron masses so they
return to free masses at freeze out.

By including the direct coupling of vector mesons to the scalar
field, we have implicitly taken into account, although partially,
the temperature dependence of hadron masses, in addition to the
density dependence.  The effective nucleon and rho-meson masses are
shown in Fig. 3 as functions of temperature, for different densities.
At 3.5$\rho _0$ and a temperature of 185 MeV, the rho meson mass reduces
to about 380 MeV.  Because of the reduction of their masses, the
abundance of the rho and omega mesons in the initial fire-cylinder
increases, and we need thus only a small pion chemical potential
of about 50 MeV in order
to reproduce the experimental pion rapidity distribution and
transverse momentum spectrum. The initial pion, rho and omega numbers are
then determined to be 44, 88, and 39, respectively.  The initial low mass
rho and omega mesons thus act as a reservoir for the observed
large pion yield.  We expect that including also
reduced in-medium masses of
mesons other than vector mesons such as the $a_1$
will further decrease the initial pion chemical potential as well as the
initial temperature.

The comparison with the experimental data for the proton and pion spectra
and rapidity distributions in this case is shown in Fig. 1 by the solid
histograms, and the agreement is again reasonable.
The final dilepton mass spectrum is shown in Fig. 2 by
the solid histogram. It is seen that with reduced rho and omega meson
masses in hot and dense hadronic matter, the agreement with the
experimental data from 2$m_\pi$ to $m_\rho$ is greatly improved. In
particular, we have about a factor of 5 enhancement in the yield of
dileptons with masses from 250 MeV to about 500 MeV.  This magnitude
is similar to that found in Ref. \cite{li95} for heavy ion collisions
at the SIS/GSI energies.  Also, the shape of the mass spectrum is now
in better agreement with the experimental data. The enhancement at low
invariant masses is mainly due to pion-pion annihilation occurring in
the hot and dense matter. Since pions have a thermal distribution,
most pion pairs are of low invariant mass. When the rho-meson mass
is reduced in hot and dense matter, its formation probability
from the pion-pion annihilation is enhanced, and thus increasing the
production of low-mass dileptons. Furthermore, we also have a good agreement
with the data around $m_{\rho ,\omega}$, while in the previous calculation
(the dotted histogram in Fig. 2) the data are overestimated by about
a factor of three. This is due to the shift of the rho-meson yield to
lower masses. The remaining peak around $m_{\rho ,\omega}$ then comes
from the decay of omega mesons which have a very small decay width,
and therefore mostly decay in the final stage when their masses have
returned to their free values.

We note that the first three points in the data
are mainly from the Dalitz decays
of $\pi_0$ and $\eta$ as pointed out in Ref. \cite{ceres95},
which are thus not essential for present discussions.
For the fourth experimental point at 250 MeV,
the Dalitz decays also contribute about half the strength
\cite{ceres95}, while according to our model another half comes
from the decay of very low mass rho mesons.

For dileptons at larger pseudo-rapidities ($3.7<\eta < 5.5$) as measured
in the HELIOS-3 collaboration \cite{helios95},
the enhancement of low mass dileptons
due to dropping vector meson masses is found to be
only about a factor of two
and is consistent with the experimental observation.
This can be understood as follows.
The dileptons measured in the CERES experiment is in the central
rapidity corresponding to the peak of the initial distribution.
These dileptons are therefore produced early in the expansion when the
baryon density is high and the rho meson mass is small.
In the HELIOS-3 experiment, the dileptons are measured in the forward
rapidity for which there are initially very few particles. They
are thus produced later in the expansion when interactions lead to a
broader rapidity distribution. The baryon density at which
large rapidity dileptons are produced is thus lower and
the reduction of the rho meson mass is smaller.

In conclusion, using
the relativistic transport model to describe the expansion stage of
a fire-cylinder formed in heavy-ion collisions at SPS energies,
we have calculated the dilepton spectra including
the medium modification of vector meson properties in hot dense matter.
It has been found that the model can explain quantitatively the observed
enhancement of
low-mass dileptons in central S+Au collisions at 200 GeV/nucleon by the
CERES collaboration.  The medium effects are expected to be more prominent
in future experiments for Pb+Pb collisions at the SPS/CERN due to the
larger initial baryon density in the collisions.

\vskip 1cm
We are grateful to Peter Braun-Munzinger, Volker Koch,
Mannque Rho, Edward Shuryak,
and Chungsik Song for helpful discussions.  We also thank
M. Murray for providing us with the preliminary data from the NA44
collaboration and for discussions.  GQL and CMK are supported by the
National Science Foundation under Grant No. PHY-9212209 and by the
Welch Foundation under Grant No. A-1110. GEB is supported by the
Department of Energy under contract No. DE-FG02-88ER40388.

\pagebreak

\centerline{\bf Figure Captions}
\begin{description}

\item {Fig. 1:} Proton and pion transverse mass spectra and rapidity
distributions. Dotted and solid histograms are obtained from simulations
based on the relativistic transport model
with free and in-medium vector meson
masses, respectively, and solid circles are
experimental data from the NA44 collaboration \cite{na44}.

\item {Fig. 2:} Dilepton invariant mass spectra from
calculations with free (dotted histogram) and in-medium vector meson
(solid histogram) masses, respectively.  Solid circles are the
experimental data from the CERES collaboration \cite{ceres95}.

\item {Fig. 3:} Effective masses of the nucleon and rho-meson as
functions of temperature for different baryon densities.

\end{description}

\end{document}